\documentclass[twocolumn,showpacs,preprintnumbers,amsmath,amssymb,eqsecnum]{revtex4}
\usepackage
{color,graphicx,mathrsfs,amsfonts}

\newcommand{\ma}[1]{\mbox{$\mathcal{#1}$}}

\begin{document}
\title{
Effects of Lovelock terms on the final fate of gravitational collapse:\\
analysis in dimensionally continued gravity}

\author{Masato Nozawa$^{1}$}
\email{nozawa@gravity.phys.waseda.ac.jp}
\author{Hideki Maeda$^{2}$}
\email{hideki@gravity.phys.waseda.ac.jp}
\address{$^{1}$Department of Physics, Waseda University, 
3-4-1 Okubo, Shinjuku-ku, Tokyo 169-8555, Japan~}
\address{$^2$ Advanced Research Institute 
for Science and Engineering,
Waseda  University, Shinjuku, Tokyo 169-8555, Japan~}

\date{\today}


\begin{abstract}
We find an exact solution in dimensionally continued gravity
in arbitrary dimensions which describes the gravitational collapse 
of a null dust fluid. 
Considering the situation that a null dust fluid 
injects into the initially anti-de Sitter spacetime,
we show that a naked singularity can be formed. 
In even dimensions, a massless ingoing null naked singularity emerges.
In odd dimensions, meanwhile, a massive timelike naked singularity forms.
These naked singularities can be globally naked
if the ingoing null dust fluid is switched off at a finite time; 
the resulting spacetime is static and asymptotically anti-de Sitter spacetime.
The curvature strength of the massive timelike 
naked singularity in odd dimensions is independent
of the spacetime dimensions or the power of the mass function.
This is a characteristic feature in Lovelock gravity.
\end{abstract}

 
\maketitle


\section{Introduction}
Black holes are one of the most fascinating objects in 
general relativity. 
It is considered that they are formed by the gravitational 
collapse of very massive stars or density fluctuations 
in the very early universe.
It has been shown that spacetimes necessarily have 
a singularity under physically reasonable conditions \cite{HE}.
Gravitational collapse is one of the presumable scenarios 
that singularities are formed.

Since there is no way to predict information 
from singularities, we cannot say anything about the 
causal future of singularities.
In order for spacetimes not to be pathological,
Penrose made a celebrated proposal, the
{\it cosmic censorship hypothesis} (CCH)~\cite{penrose}, 
that is, singularities which are formed in a 
physically reasonable gravitational collapse 
should not be seen by distant observers, i.e., 
spacetimes are asymptotically predictable.
(See~\cite{HE}.)
This is the weak version of the CCH.
The strong version of the CCH states that 
no observer can see the singularities 
formed by gravitational collapse~\cite{penrose1979}. 
Although there is a long history of research 
on the final fate of the gravitational collapse
\cite{ns-dust,CMP,ns-vaidya,ns-pf,ns-scalar,lake1992,
ns-nonspherical,ns-others,ns-numerical}, 
we are far from having achieved consensus on the validity of the CCH, 
which is one of the most important open problems in general relativity.
(See~\cite{harada} for a recent review.)

Lately it has been of great importance to consider 
higher-dimensional spacetimes. Although there is no
direct observational evidence of extra dimensions,
fundamental theories such as string/M-theory predict 
the existence of extra dimensions.
Over the past years, the braneworld model has also reinforced
the study of higher-dimensional spacetimes 
\cite{LED}.

There exists a natural extension of general relativity
in higher dimensions, Lovelock gravity~\cite{lovelock}.
Lovelock Lagrangian is composed of dimensionally 
extended Euler densities, which include the higher-order 
curvature invariants with a special combination 
so that the field equations are of second order \cite{zumino}. 
In four dimensions, such higher curvature terms
do not contribute to the Einstein field equations
since they culminate in total derivatives.
The quadratic curvature terms are known as Gauss-Bonnet terms,
which appear in the low energy limit of 
heterotic string theory~\cite{GB}. 
In Gauss-Bonnet gravity, static black hole solutions 
have been found and investigated in detail
\cite{GBBH,GBstatic}.

Higher-order curvature terms come into effect where gravity becomes very strong.
One of the present authors has discussed the gravitational 
collapse of a null dust fluid in Gauss-Bonnet gravity~\cite{hideki}.
It was found that the final fate of the 
gravitational collapse in five dimensions is
significantly different from that in general relativity:
a massive timelike naked singularity can be formed.
Such a class of singularities has never appeared 
in the general relativistic case~\cite{lake1992}.
The purpose of the present paper is to analyze how higher-order Lovelock terms 
modify the final fate of gravitational collapse
in comparison to the Gauss-Bonnet or general relativistic cases.

A static vacuum solution in Lovelock gravity has been found
\cite{whitt,Cai} and studied~\cite{myers}. 
In this case, the metric function is obtained from an algebraic equation, 
which cannot be solved explicitly in general.
This is because, in $D$-dimensional spacetimes,
Lovelock Lagrangian contains $[(D+1)/2]$ independent free parameters,
where the symbol $[x]$ is understood to take the integer part of $x$. 
This complexity makes the analysis in Lovelock gravity
quite hard to tackle, and hence it is not obvious
to extract physical information from the solution.
Consequently, Ba\~nados, Teitelboim and Zanelli have proposed 
a method that reduces the [$(D+1)/2$] free parameters 
to only two by taking a special relation 
between the coefficients of Lovelock Lagrangian~\cite{BTZ2}. 
An exact solution in the theory of gravity, 
called dimensionally continued gravity (DC gravity), 
representing a static vacuum solution, has been found~\cite{BTZ2,cai}.
DC gravity is very effective in estimating 
the effects of higher-order Lovelock terms 
although it is a restricted class of Lovelock gravity.

In this paper, we extend our investigations on the 
gravitational collapse of a null dust fluid 
in Gauss-Bonnet gravity~\cite{hideki}
into DC gravity in order to see the effects 
of higher-order Lovelock terms on the final fate of 
gravitational collapse.
An exact solution describing gravitational collapse 
of a null dust fluid is given. 
We discuss the final fate of gravitational collapse
of a null dust fluid
in comparison with the results in 
Gauss-Bonnet gravity~\cite{hideki}.


The outline of this paper is as follows;
in Sec. \ref{sec:LG}, 
DC gravity is summarized.
An exact solution for a null dust fluid is presented in 
Sec. \ref{sec:NDS}. Sec. \ref{sec:NSF} is devoted to
the discussion of naked singularity formations
in the gravitational collapse of a null dust fluid.
The strength of naked singularities is investigated
in Sec. \ref{sec:strength}.
Our conclusions and discussion follow in
Sec. \ref{sec:conclusions}.
The discussions on the geodesics from singularities
are given in Appendix \ref{sec:geodesics}.
In Appendix \ref{sec:sc}, 
the scalar curvature quantities are
calculated for the estimation of 
singular nature. We discuss the general relativistic case in Appendix
\ref{sec:gr} so as to compare it with the DC gravity results. 
We use the unit conventions $c=1$ throughout the paper.


\section{Dimensionally continued gravity}
\label{sec:LG}
We first summarize DC gravity as a restricted class of Lovelock gravity.
Then we give the field equations in DC gravity
for odd and even dimensions.

Lovelock gravity is a natural extension of
general relativity in $D (\geq 3)$-dimensional spacetimes,
whose action is given by~\cite{lovelock}
%
\begin{align}
I=\int \mathcal L_D +I_m 
\label{eq:lovelock1}
\end{align}
with
\begin{align}
\int \mathcal L_D&\equiv \kappa 
\sum _{p=0}^{[D/2]}\alpha _p I_p,
\label{eq:lag} \\
I_p&\equiv \int _{\mathcal M}
{\epsilon }_{a _1\cdots a_D}
\mathcal R^{a _1a _2} \wedge \cdots \wedge
\mathcal R^{a _{2p-1}a _{2p}}\nonumber \\
&~~~~~~~~~~~~~~~~~~~~~~~~\wedge
e^{a _{2p+1}}\wedge \cdots \wedge
e^{a _D},
\label{eq:lovelock2}
\end{align}
where 
$\ma R^{ab}=d\omega ^{ab}+{\omega^a}_c\wedge {\omega ^c}_b$
is the curvature two-form, $e^a$  is the orthonormal
frame, $\omega ^{ab}$ is the spin connection
which satisfies the torsion-free Cartan structure equations
and $I_m$ is the action of the matter field, respectively.
In our notation, $a_p, b_p, \cdots ,$ 
runs over all spacetime dimensions
$0,\cdots , D-1$. 
$\epsilon _{a_1\cdots a_D}$ is the $D$-dimensional 
Levi-Civita tensor in Minkowski space 
($\epsilon _{01\cdots (D-1)}=1$).
The [$(D+1)/2$] real constants $\alpha _p$ in the action
(\ref{eq:lag}) have dimensions [length]${}^{-(D-2p)}$; 
otherwise they are arbitrary.
A real constant $\kappa $ has units of the action and 
can be set to be positive without loss of generality.
$(\kappa \alpha _1)^{-1}$ is proportional to the 
$D$-dimensional gravitational constant 
with a positive coefficient, 
so that it is real and positive.

The Lovelock action (\ref{eq:lovelock1}) is the special 
combination of curvature invariants in that 
the resulting field equations do not include 
more than the second derivatives of the metric.
They are composed of dimensionally extended Euler densities
\cite{zumino}. In $D$-dimensions, the first 
[$(D+1)/2$] terms contribute to the field equations;
higher-order curvature terms will be the total derivatives. 
In even dimensions, the last term in the summation
(\ref{eq:lovelock1}), that is $p=D/2$, 
is the Euler density and consequently does not contribute to
the field equations. 
The action (\ref{eq:lovelock1}) reduces to the 
usual Einstein-Hilbert action in four dimensions.

Varying the action (\ref{eq:lovelock1}) with respect to
the vielbein form $e^a$, one obtains the field equations;
\begin{align}
&-\kappa \sum _{p=0}^{[D/2]}\alpha _p(D-2p)
{\epsilon }_{a _1\cdots a_D}
\mathcal R^{a _1a _2}\wedge \cdots \nonumber \\
& \qquad  \wedge
\mathcal R^{a _{2p-1}a _{2p}}\wedge
e^{a _{2p+1}}\wedge \cdots \wedge
e^{a _{D-1}}=Q_{a_D},
\label{eq:fieldeq1}
\end{align}
where $Q_{a_D}$ is the $(D-1)$-form representing
the energy momentum tensor $T_{\mu \nu }$ of
the matter,
\begin{align}
Q_{a_D}\equiv \frac{1}{(D-1)!}{T_{a_D}}^{b_1}
{\epsilon }_{b _1\cdots b_D}
e^{b_2}\wedge \cdots \wedge e^{b_D}.
\label{eq:mattertensor1}
\end{align}
The spherically symmetric vacuum solution
in Lovelock gravity was first found by
Whitt~\cite{whitt}.
This solution has been extended to spacetime as a product manifold
$\ma M \approx \ma M^2\times \ma K^{D-2}$ by Cai~\cite{Cai},
where $\ma K^{D-2}$ is the maximally symmetric 
spaces with constant curvature $(D-2)(D-3)k$.
Without loss of generality, $k$ is normalized as
$+1$ (positive curvature), $0$ (zero curvature) 
and $-1$ (negative curvature), respectively.


The vacuum static solution 
in Lovelock gravity includes
$[(D+1)/2]$ arbitrary constants~\cite{whitt,Cai},
so that the solution
are given by the roots of polynomial of
$[(D-1)/2]$ degrees,
which cannot be written down in explicit form.
In DC gravity, the arbitrary constants are reduced 
to two by embedding the Lorentz group
$SO(D-1, 1)$ into the larger AdS group 
$SO(D-1, 2)$~\cite{BTZ2}.
The remaining two arbitrary constants are 
a gravitational constant and a cosmological constant.
Accordingly, Lovelock gravity is separated into 
two distinct type of branches for odd and
even dimensions.
The special combinations of Lovelock coefficients
are given by
\begin{align}
\alpha _p =\left\{
\begin{array}{ll}
\dfrac{1}{D-2p} 
\left(
\begin{array}{cc}
n-1 \\
p
\end{array}
\right)l^{-D+2p}, \quad & \textrm{for $D=2n-1$},
\\
\\
\left(
\begin{array}{cc}
n \\
p
\end{array}
\right)l^{-D+2p}, & \textrm{for $D=2n$},
\end{array}
\right.
\label{eq:coefficient}
\end{align}
where $n\equiv [(D+1)/2]$ and
$1/l^2$ is proportional to the cosmological constant with 
a negative coefficient.
From Eq.~(\ref{eq:coefficient}), we obtain
\begin{align}
(\kappa\alpha _1)^{-1} =\left\{
\begin{array}{ll}
(D-2)/[(n-1)\kappa l^{-D+2}], \quad & 
\textrm{for $D=2n-1$},
\\
\\
1/(n\kappa l^{-D+2}), & 
\textrm{for $D=2n$},
\end{array}
\right.
\label{eq:G-Ddim}
\end{align}
so that $l$ must be real and positive both 
for $D=2n-1$ and $D=2n$ 
in order for the $D$-dimensional gravitational constant 
to be real and positive.

In odd dimensions $D=2n-1$, Lagrangian 
$\ma L_{2n-1}$ is given by~\cite{BTZ2}
\begin{align}
\ma L_{2n-1}=&\kappa \sum_{p=0}^{n-1}\alpha _p
{\epsilon }_{a _1\cdots a_{2n-1}}
\ma R^{a_1a_2}\nonumber \\
&\wedge \cdots \wedge
\ma R^{a _{2p-1}a _{2p}} \wedge
e^{a _{2p+1}}\wedge \cdots \wedge
e^{a _{2n-1}}
\label{eq:oddlaglange}
\end{align}
with (\ref{eq:coefficient}).
We adopt units such that
\begin{align}
 {(D-2)!\Omega _{D-2}}\kappa l^{-1}=1
\qquad \textrm{for $D=2n-1$},
\label{eq:oddunit}
\end{align}
where $\Omega _{D-2}$ is the area of 
$(D-2)$-dimensional unit sphere
\begin{align}
\Omega_{D-2}=\frac{2\pi^{(D-1)/2}}{\Gamma ((D-1)/2)}.
\end{align}
Eq. (\ref{eq:oddlaglange}) is a $D$-dimensional Chern-Simons Lagrangian
for AdS group, whose exterior derivative is proportional to
the Euler density in $(D+1)$-dimensions.
Defining two-forms
\begin{align}
\hat{\ma R}^{ab}\equiv \ma R^{ab}+l^{-2}e^a\wedge e^b,
\end{align}
one can find the equations of motion for
odd dimensions $D=2n-1$,
\begin{align}
-l^{-1}\kappa {\epsilon }_{a _1\cdots a_{2n-1}}
\hat{\ma R}^{a_1a_2}\wedge \cdots \wedge
\hat{\ma R}^{a _{2n-3}a _{2n-2}} 
=Q_{a_{2n-1}}.
\label{eq:oddfieldeq}
\end{align}

%
%
In even dimensions $D=2n$, Lagrangian is 
given by
\begin{align}
\ma L_{2n}=\kappa 
{\epsilon }_{a _1\cdots a_{2n}}
\hat{\ma R}^{a_1a_2}\wedge \cdots \wedge
\hat{\ma R}^{a _{2n-1}a _{2n}},
\label{eq:evenlaglange}
\end{align} 
where we adopt the units as
\begin{align}
{2D(D-2)!\Omega _{D-2}}\kappa l^{-2}=1
\qquad \textrm{for $D=2n$}.
\label{eq:evenunit}
\end{align}
Due to the particular choice of the Lovelock
coefficients (\ref{eq:coefficient}),
the Lagrangian (\ref{eq:evenlaglange}) is 
brought into the Born-Infeld type
\begin{align}
\ma L_{2n}=\kappa \sqrt{|{\rm det}\hat{\ma R}^{a_1a_2}|}.
\end{align}
The field equations for even dimensions $D=2n$
are then given by
\begin{align}
-2nl^{-2}\kappa {\epsilon }_{a _1\cdots a_{2n}}
\hat{\ma R}^{a_1a_2}\wedge \cdots &\nonumber \\
\wedge \hat{\ma R}^{a _{2n-3}a _{2n-2}}&
\wedge e^{a_{2n-1}} 
=Q_{a_{2n}}.
\label{eq:evenfieldeq}
\end{align}


\section{Null dust solution}
\label{sec:NDS}
%
Ba\~nados, Teitelboim and Zanelli derived 
the static and spherically symmetric vacuum
solution of DC gravity in 
$D (\geq 3)$-dimensions~\cite{BTZ2}. 
The extension to spacetime 
$\ma M \approx \ma M^2\times \ma K^{D-2}$
has also been considered \cite{cai}.
The general metric on the spacetime $\ma M$ can be written as
\begin{align}
g_{\mu\nu}=\mbox{diag}(g_{AB},r^2\gamma_{ij}),
\label{eq:structure}
\end{align}
where $g_{AB}$ is an arbitrary Lorentz metric on $\ma M^2$, 
$r$ is a scalar function on $\ma M^2$ 
with $r=0$ defining the boundary of $\ma M^2$ and 
$\gamma_{ij}$ is the unit curvature metric on $\ma K^{D-2}$. 
When $r$ is not constant, it is shown 
in full Lovelock gravity that 
the vacuum solutions are described by the 
static solutions obtained by Cai~\cite{Cai}; i.e,
the generalization of the Birkhoff's theorem 
into Lovelock gravity holds~\cite{zegers}.

From this fact, when $r$ is not constant, 
the general vacuum solution
in DC gravity is obtained as
~\cite{BTZ2,cai}
\begin{align}
ds^2=-f(r)dt^2+f^{-1}(r)dr^2+r^2d\Sigma _{k,D-2}^2
\label{eq:exsol}
\end{align}
with
\begin{align}
f(r)=\left\{
\begin{array}{ll}
k-(2M/r)^{1/(n-1)}+l^{-2}r^2, \quad &
{\rm for}\enspace D=2n, \\ \\
k-M^{1/(n-1)}+l^{-2}r^2, \quad &
{\rm for}\enspace D=2n-1,
\end{array}
\right.
\label{eq:fr}
\end{align}
where we denote the line element of 
$\ma K^{D-2}$ as $d\Sigma _{k,D-2}^2$.

It is noted that the conventions of $M$ are slightly 
different from those in~\cite{BTZ2,cai}.
The constant $M$ has dimensions of length
in even dimensions, while it is dimensionless 
in odd dimensions. This is owing to the definition
of the present unit conventions.
(See Eqs. (\ref{eq:oddunit}) and (\ref{eq:evenunit}).) 
The solution (\ref{eq:exsol}) is asymptotically  
$D$-dimensional AdS spacetime because $l^2>0$.

The causal structures in the spherically 
symmetric case ($k=1$) are shown in~\cite{BTZ2}.
The solution represents a black hole for $M>0$, 
and represents the $D$-dimensional AdS spacetime 
for $M=0$ in even dimensions.
In odd dimensions, meanwhile, the solution represents 
a black hole for $M^{1/(n-1)}>k$,
the $D$-dimensional AdS spacetime for $M=0$
and the naked singularity for $M^{1/(n-1)}<k$.
In three dimensions, this is the so-called BTZ solution 
~\cite{BTZ,bhtz}, in which there are 
no curvature singularities.
In four dimensions with $k=1$, 
Eq. (\ref{eq:exsol}) 
reduces to the Schwarzschild-AdS solution.

We will see in five and six dimensions, 
the solution (\ref{eq:fr}) belongs to the 
special family of solution in Gauss-Bonnet gravity.
In five and six dimensions, up to quadratic curvature terms 
have nontrivial effects on the field equations.
The Lagrangian in Gauss-Bonnet gravity is 
\begin{align}
L &=-2\Lambda +R+\alpha 
(R^2 -4R_{\mu \nu }R^{\mu \nu }+
R_{\mu \nu \rho \sigma }R^{\mu \nu \rho \sigma }).
\end{align}
The vacuum static solution in Gauss-Bonnet gravity \cite{GBBH}
is given by the metric (\ref{eq:exsol}) with
\begin{align}
f= k+\frac{r^2}{2\tilde{\alpha }}\left[1-
\sqrt{1+4\tilde{\alpha }\left(
\frac{m}{r^{D-1}}+\tilde{\Lambda }\right)}\right],
\label{eq:GBBH}
\end{align}
where $\tilde{\alpha }\equiv (D-3)(D-4)\alpha $,
$\tilde{\Lambda }\equiv 2\Lambda /[(D-1)(D-2)]$ and
$m$ is the integration constant.
We can find easily the solution (\ref{eq:GBBH}) 
reduces to the solution (\ref{eq:fr}) 
with $1+4\tilde{\alpha }\tilde{\Lambda }=0$ and
$2\tilde{\alpha }=l^2$ in five and six dimensions.

In four-dimensional spacetimes,
we can obtain the Vaidya solution 
by replacing the mass parameter $M$
in the Schwarzschild solution 
as an arbitrary mass function $M(v)$,
which specifies the flux of the ingoing radiation
with the advanced time coordinate $v$.
In the course of this heuristic procedure,
we find an exact solution of DC gravity 
in $D$-dimensional spacetimes
by taking $M\to M(v)$ in the solution (\ref{eq:exsol}),
representing a radially ingoing null dust fluid.
The energy momentum tensor 
of a null dust fluid is written as
%
\begin{align}
T_{\mu \nu }=\rho (v, r)l_\mu l_\nu ,
\label{eq:setensor}
\end{align}
where $\rho (v, r)$ represents the energy density
of the null dust fluid 
and  $l^\mu $ is the null vector 
normalized as $l^\mu \partial _\mu =-\partial _r$.
%
Then we can find the metric of the null dust 
solution as
\begin{align}
ds^2=-f(v, r)dv^2+2dvdr+r^2d\Sigma _{k, D-2}^2
\label{eq:sol}
\end{align}
with
\begin{widetext}
\begin{align}
f(v, r)=\left\{
\begin{array}{ll}
k-(2M(v)/r)^{1/(n-1)}+l^{-2}r^2,\qquad  & 
{\rm for}\enspace D=2n, \\ \\
k-M(v)^{1/(n-1)}+l^{-2}r^2,  &
{\rm for}\enspace D=2n-1.
\end{array}
\right.
\label{eq:fvr}
\end{align}
\end{widetext}
We can confirm the solution (\ref{eq:sol}) 
to satisfy the field equations with the help of
\begin{align}
 \hat{\ma R}^{vr}&=\frac{n}{2(n-1)^2r^2}
\left(\frac{2M(v)}{r}\right)^{1/(n-1)}
e^v\wedge e^r,
\\
\hat{\ma R}^{ri}&=-\frac{1}{2(n-1)r^2}
\left(\frac{2M(v)}{r}\right)^{1/(n-1)} 
e^r\wedge e^{i} 
\nonumber \\
&~~+\frac{\dot{M}(v)}{2(n-1)rM(v)}
\left(\frac{2M(v)}{r}\right)^{1/(n-1)}
e^v\wedge e^{i},
\\
\hat {\ma R}^{vi}&=-\frac{1}{2(n-1)r^2}
\left(\frac{2M(v)}{r}\right)^{1/(n-1)}
e^v\wedge e^{i},
\\
\hat{\ma R}^{ij}&=\frac{1}{r^2}
\left(\frac{2M(v)}{r}\right)^{1/(n-1)}
e^{i}\wedge e^{j},
\end{align}
for even dimensions and
\begin{align}
\hat{\ma R}^{ri}&=\frac{\dot{M}(v)
M(v)^{-1+1/(n-1)}}{2(n-1)r}
e^v\wedge e^{i},
\\
\hat{\ma R}^{ij}&=\frac{M(v)^{1/(n-1)}}{r^2}
e^{i}\wedge e^{j},
\end{align}
for odd dimensions, where the overdot
denotes the differentiation with respect to
the coordinate $v$, and $i, j$ represent the indices
on $\ma K^{D-2}$.
The matter terms are given by
\begin{align}
Q_v=\frac{\rho (v,r)}{(D-2)!}
{\epsilon }_{rva_3\cdots a_{D}}
e^v\wedge e^{a_3}\wedge \cdots \wedge e^{a_D}.
\end{align}
From the field equations,
the energy density of the null 
dust fluid is given by
\begin{align}
\rho (v, r)=\frac{1}{\Omega _{D-2}r^{D-2}}\dot{M}
\label{eq:ED}
\end{align}
both in odd and even dimensions.
$\dot{M}\geq 0$ is required in order for 
the energy density 
of the null dust fluid to be non-negative.
The solution (\ref{eq:sol}) is a generalization 
of the Vaidya solution. 
Eq. (\ref{eq:sol}) reduces to Eq. (\ref{eq:exsol})
with $dv=dt+dr/f(r)$
when the mass function $M$ is constant.


\section{Naked Singularity Formations}
\label{sec:NSF}
In this and the next sections, we study the final fate of the gravitational collapse 
of a null dust fluid in DC gravity by use of the solution obtained 
in the previous section.
In what follows, we restrict attention to the spherically symmetric case, 
i.e., $k=1$, for comparison with the result 
in Gauss-Bonnet gravity~\cite{hideki}.
Hereafter, we shall call the solution (\ref{eq:sol}) 
with $k=1$ the DC-Vaidya solution.
For the case of constant $M$,
we shall call the solution the DC-BTZ solution.
In the Vaidya solution, the mass function $M(v)$ is the Misner-Sharp 
mass.
In the generalized Vaidya solution in Gauss-Bonnet gravity, $M(v)$ is 
not the Misner-Sharp mass but more preferable quasi-local mass
\cite{hideki}. 
In this paper, we also adopt $M(v)$ as the quasi-local mass 
in the DC-Vaidya solution.
The odd-and even-dimensional cases are
separately investigated below.

Discussions with the fixed point method are also presented in Appendix \ref{sec:CMP}.


\subsection{Even dimensions}
We consider the situation in which a null dust fluid 
radially falls into the initial AdS spacetime ($M(v)=0$)
at $v=0$. 
We set the mass function as a power law form
\begin{align}
M(v)=M_0v^q
\end{align}
for simplicity, where $M_0$($>0$) and $q (\geq 1)$ 
are constant. 
We can see easily from Eq. (\ref{eq:ED}) that
a central singularity appears at $r=0$ for $v>0$.
Let us discuss more specifically the singular nature of
$v=r=0$.

The future-directed outgoing radial null geodesics
obey 
\begin{align}
\frac{dr}{dv}=\frac{1}{2}f,
\label{eq:null}
\end{align}
so a trapped region exists in $f\leq 0$,
which is given by
\begin{align}
2M(v)=2M_0v^{q} \geq r(1+l^{-2}r^2)^{n-1},
\label{eq:AHeven}
\end{align}
where the equality sign holds at the 
marginally trapped surface $f=0$.
Thus the singularity at $r=0,~v>0$ 
is in the trapped region and only the 
point $v=r=0 $ has the possibility of being naked.
Along the trapping horizon, that is, 
the trajectory of the marginally 
trapped surface, we have
\begin{align}
 ds^2=\frac{4qM_0v^{q-1}}{(1+l^{-2}r^2)^{n-2}
(1+(2n-1)l^{-2}r^2)}dv^2,
\end{align}
so that it is spacelike for $v>0, r=0$.

We try to find the radial null geodesics emanating 
from the singularity in order to determine whether
or not the singularity is naked.

It is shown that if a future-directed radial null geodesic 
does not emanate from the singularity, nor does a future-directed causal 
(excluding radial null) geodesic. 
(See Appendix~\ref{sec:nullgeodesics} for the proof.)
So we consider here only the future-directed outgoing radial null geodesics.

Suppose the asymptotic form of the geodesics 
as $v\simeq K_1r^p $ around $v=r=0$ 
~\cite{joshibook},
where $p$ and $K_1$ are positive constants. 
Then the lowest order of Eq. (\ref{eq:null})
around $v=r=0$ yields $p=1$, so that 
\begin{align}
v\simeq K_1r.
\label{eq:asyeven}
\end{align}
$K_1=2$ is obtained for $q>1$, while
$K_1$ satisfies the relation
\begin{align}
\left(1-\frac{2}{K_1}\right)^{n-1}=2M_0K_1
\label{eq:keven}
\end{align}
for $q=1$.
The condition for the existence of a positive 
$K_1$ satisfying Eq. (\ref{eq:keven}) 
is 
\begin{align}
0<M_0\leq \frac{1}{4n}\left(
\frac{n-1}{n}\right)^{n-1},
\label{eq:M_0}
\end{align}
and then $K_1> 2$ holds.

Along the radial null geodesic arising from
$v=r=0$ with asymptotic form (\ref{eq:asyeven}),
the energy density for the null dust fluid 
(\ref{eq:ED}) and the Kretschmann scalar 
$\ma I_1=R_{\mu \nu \rho \sigma }R^{\mu \nu \rho \sigma }$ 
diverge for $r\to 0$ as (see Appendix \ref{sec:sc})
\begin{align}
\rho &=\ma O(1/r^{D-q-1})
\end{align}
and
\begin{align}
\ma I_1&=\ma O(1/r^{4(1-(q-1)/(D-2))}),
\end{align}
respectively, so they are singular null geodesics
for $1\leq q<D-1$.
Thus, the spacetime represents the formation
of a naked singularity for $1< q<D-1$ with any $M_0(>0)$ 
and for $q=1$ with $M_0$ satisfying the condition (\ref{eq:M_0}).
In the limit of $n\to \infty $,
the right hand side of Eq. (\ref{eq:M_0})
goes to zero; i.e., the formation of 
naked singularity is less plausible 
as the spacetime dimensions are higher.
Hereafter we consider the case where $1\leq q<D-1$,
in which a naked singularity appears at $v=r=0$.

The central singularity $v=r=0$ is
then at least locally naked. 
Now we consider the structure of the naked singularity.
Expanding Eq.~(\ref{eq:AHeven}) around $r=0$, 
we have
\begin{eqnarray}
v \simeq(2M_0)^{-1/q}r^{1/q}.
\end{eqnarray}  
For $q>1$, there exists a spacetime region ${\cal U}$ which is both the past of the trapping horizon and the future of the future-directed outgoing radial null geodesic $\gamma$ which behaves as Eq.~(\ref{eq:asyeven}) near $v=r=0$.
Such a region also exists for $q=1$ because $1>2M_0K_1$ holds from Eq.~(\ref{eq:keven}) and $K_1>2$.
Because the trapping horizon is spacelike for $v>0$ and $r>0$, the past-directed ingoing radial null geodesic $\zeta$ emanating from an event in ${\cal U}$ never crosses the trapping horizon.
Also $\zeta$ never crosses $\gamma$ except for $v=r=0$ because they are both future-directed outgoing radial null geodesics.
Consequently, $\zeta$ inevitably reaches the singularity at $v=r=0$.
Since ${\cal U}$ is an open set, we then conclude that there exist an infinite number of future-directed
outgoing null geodesics emanating from the singularity at $v=r=0$.
Such geodesics should correspond to the solution of 
Eq.~(\ref{eq:geodesictheta}) with $\eta=0$ or 
$\eta=\infty$ in Appendix \ref{sec:CMP}.
On the other hand, the future-directed ingoing radial null geodesic 
reaching the singularity $v=r=0$ is only $v=0$. 
Therefore, it is concluded that the singularity $v=r=0$ 
has an ingoing-null structure.

When $M_0$ fails to satisfy the condition
(\ref{eq:M_0}) for $q=1$
in $D\geq 4$ $(n\geq 2)$ dimensions, 
more detailed analyses
are needed in order to determine whether
the final fate of gravitational collapse
is a black hole or a naked singularity
since there might exist null geodesics 
which do not obey the power-law form
asymptotically.

\begin{figure}[h]
\centerline{
\includegraphics[width=6cm]{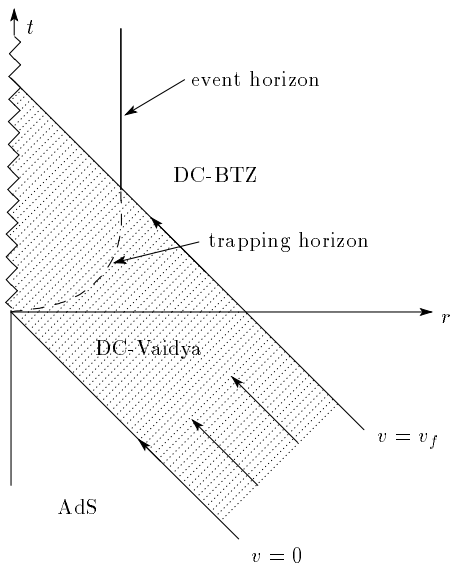}
}
\caption{Illustration of the gravitational 
collapse of a null dust fluid in even dimensions.
A central singularity forms at $v=r=0$.
The DC-BTZ solution is attached to 
$D$-dimensional AdS spacetime
via the DC-Vaidya solution. 
Zigzag lines denote the central singularity.
}
\label{fig:even}
\end{figure}

In order to see whether the singularity is 
globally naked,
we consider a very simple situation in which 
the null dust fluid is switched off at $v=v_f>0$. 
The solution is described by
the DC-BTZ solution for $v>v_f$,
which is static and asymptotically AdS spacetime. 
The DC-BTZ solution is joined with the $D$-dimensional 
AdS spacetime for $v<0$
by way of the DC-Vaidya solution for $0\leq v\leq v_f$. 
The situation is depicted in Fig. \ref{fig:even}.

\begin{figure}[h]
\centerline{
\includegraphics[width=7cm]{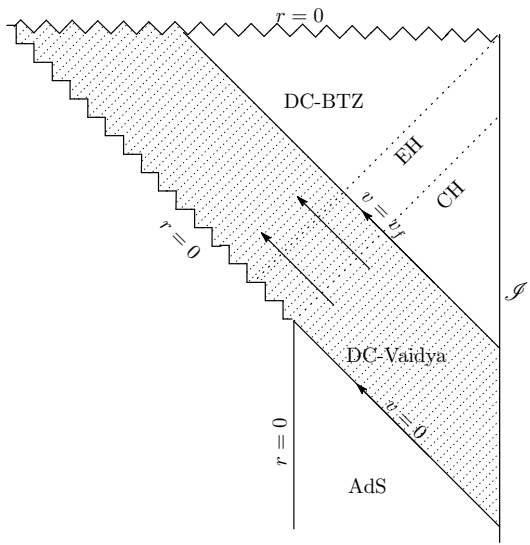}
}
\caption{
The figure shows a Penrose diagram
for a globally naked singularity formation.
Infinity consists of a timelike surface $\mathscr I$.
EH and CH denote the event horizon and 
the Cauchy horizon, respectively.
}
\label{fig:pd1}
\end{figure}

If a null geodesic emanating from
the singularity $v=r=0$ reaches the surface
$v=v_f$ in the untrapped region,
it can escape to infinity;
then, the singularity is globally naked.
If we take $v_f$ to be sufficiently small,
the singularity can be globally naked.
The singularity in the outer DC-BTZ spacetime is spacelike in even dimensions~\cite{BTZ2}.
Thus, the Penrose diagram of the gravitational collapse 
is drawn in Fig. \ref{fig:pd1} for the globally 
naked singularity formation. 
Of course, the locally naked singularity formation 
is also obtainable.



\subsection{Odd dimensions}
\label{subsec:odd}
As in the even-dimensional case, 
we consider the situation in which a null dust 
fluid radially falls into the
initial AdS spacetime, and we choose 
the mass function as the power-law form $M(v)=M_0v^q$.
Then we can see from Eq. (\ref{eq:ED}) that
a singularity develops at $r=0$ for $v> 0$.
As shown below, the point $v=r=0$ may also be singular.
It is noted that the singularity with $r=0$ and $v>0$ 
is also interpreted as a conical singularity; that is, 
a deficit angle exists at $r=0$. (See~\cite{bhtz} 
for the conical singularity in the BTZ solution.)

%
A trapped region $f\leq 0$ is given by
\begin{align}
M(v)=M_0v^q\geq (1+l^{-2}r^2)^{n-1},
\label{eq:AHodd}
\end{align}
where the equality holds at the 
marginally trapped surface $f=0$.
Along the trapping horizon, we have
\begin{align}
ds^2=\frac{M_0ql^2v^{q-1}}{r(n-1)(1+l^{-2}r^2)^{n-2}}dv^2,
\end{align}
so that it is spacelike for $r>0, v>0$.
From Eq. (\ref{eq:AHodd}), 
the singularity $r=0$ might be naked 
for $0\leq v\leq v_{AH}$, 
where $v_{AH} \equiv M_0^{-q}$. 
Following the argument in Appendix \ref{sec:nullgeodesics}, 
we consider only the future-directed outgoing
radial null geodesics from the singularity.
Suppose the asymptotic form
\begin{align}
 v\simeq v_0+K_2r^p
\label{eq:asy}
\end{align}
near $r=0$ with $0\leq v_0\leq v_{AH}$, 
where $p$ and $K_2$ are positive constants. 
From the lowest order of Eq. (\ref{eq:null}),
we obtain
\begin{align}
v\simeq v_0+\frac{2}{1-(M_0v_0^{q})^{1/(n-1)}}r
\label{eq:asyodd}
\end{align}
for $0\leq v_0< v_{AH}$, 
while there are no null geodesics with the
asymptotic form Eq. (\ref{eq:asy}) for $v=v_{AH}$.

Along the radial null geodesics 
from the singularity $v=r=0$
asymptoting to Eq. (\ref{eq:asyodd}),
the Kretschmann scalar diverges as
\begin{align}
\ma I_1& = \ma O(1/r^{4(1-q/(D-1))})
\end{align}
for $D \ge 5$.
In the meanwhile, it is finite for $D=3$.
In fact, the scalar quantities for the DC-Vaidya solution 
with $D=3$ are the same as those for the three-dimensional 
AdS spacetime, which are constant everywhere; 
$\ma I_1=\ma I_2=12l^{-4}$, $R=-6l^{-2}$. 
(The curvature invariants $\ma I_1$, $\ma I_2$ and $R$ for the 
metric (\ref{eq:sol}) are calculated in Appendix \ref{sec:sc}.)
However, the energy density of the null dust fluid (\ref{eq:ED}) 
diverges along the null geodesics (\ref{eq:asyodd}) as
\begin{align}
\rho & = \ma O(1/r^{D-q-1})
\end{align}
for $D \ge 3$.
As a result, they are singular null geodesics
for $1\leq q<D-1$. 

On the other hand, the radial null geodesics
from the singularity $r=0$, $0<v<v_{AH}$
asymptoting to Eq. (\ref{eq:asyodd}) are singular 
null geodesics for any $q (\ge 1)$.
Along them, the Kretschmann scalar and the energy density of the
null dust fluid diverge for $D \ge 5$ and $D \ge 3$, respectively.

\begin{figure}[h]
\centerline{
\includegraphics[width=8cm]{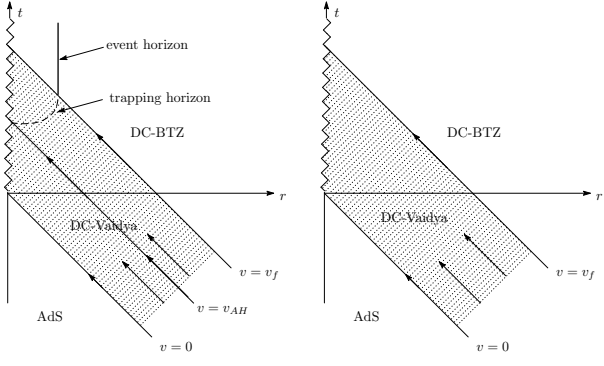}
}
\caption{Gravitationl collapse in odd dimensions 
for $v_f> v_{AH}$ (left) and $0<v_f \le v_{AH}$ (right).
The DC-Vaidya solution for $0\leq v\leq v_f$ is 
jointed with the DC-BTZ solution for $v>v_f$ and
the AdS spacetime for $v<0$.}
\label{fig:odd}
\end{figure}

We can consider the situation as in the even-dimensional 
case; that is, the null dust fluid is turned off
at some finite time $v=v_f>0$ in order to see
whether the singularity is globally naked.
Fig. \ref{fig:odd} illustrates the situation in which
the DC-BTZ solution for $v>v_f$ is connected
with the $D$-dimensional AdS spacetime for $v<0$ 
via the DC-Vaidya solution for $0\leq v\leq v_f$.

The structures of the singularity in the outer DC-BTZ spacetime 
in odd dimensions are spacelike, null and timelike for 
$v_{\rm f}>v_{\rm AH}$, $v_{\rm f}=v_{\rm AH}$ and 
$v_{\rm f}<v_{\rm AH}$, respectively \cite{BTZ2}.
Now we consider the structure of the naked singularity 
in the null dust region.
The naked singularity with $0 \le v<v_{AH}$ is both 
an endpoint of the future-directed ingoing radial null geodesic 
and the initial point of the future-directed outgoing radial 
null geodesic, therefore we conclude that it is timelike.

For $0<v_f\leq v_{AH}$, the singularity is
always globally naked.
In the case of $v_f>v_{AH}$, 
if we take $v_{\rm f}$ as sufficiently close to $v_{AH}$, 
a null ray emanating from the singularity reaches 
the surface $v=v_{f}$ in the untrapped region, so that 
it can escape to infinity;
then, the singularity is globally naked.
As in the even-dimensional case, 
globally naked singularities can be formed
in odd dimensions.

\begin{figure}[h]
\begin{center}
\includegraphics[width=4.5cm]{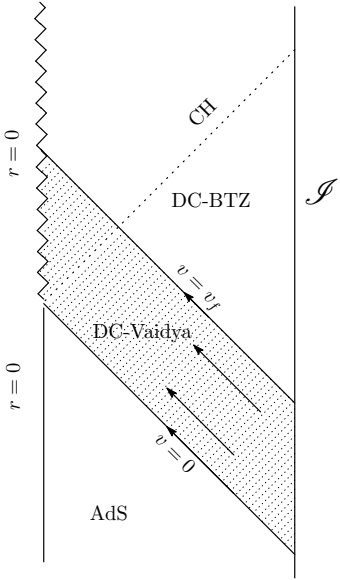}
\end{center}
\caption{Penrose diagram in odd dimensions 
for $0< v_f<v_{AH}$. The massive timelike 
naked singularity appears. The resultant
spacetime is always globally naked.}
\label{fig:oddpd}
\end{figure}

\begin{figure}[h]
\centerline{
\includegraphics[width=8cm]{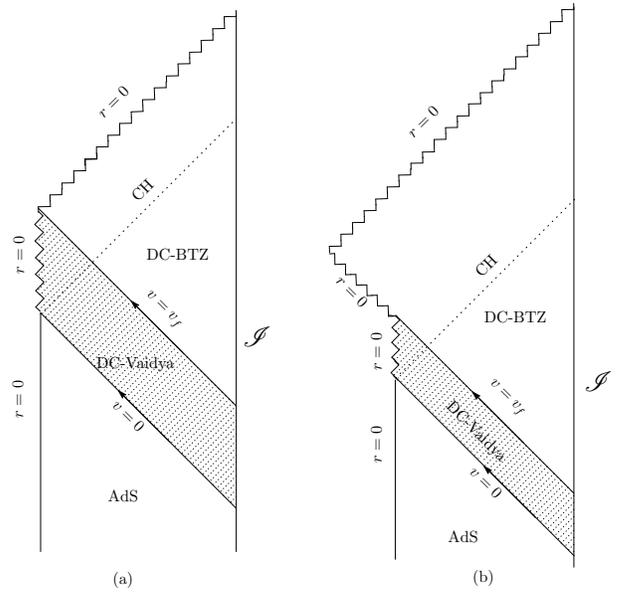}
}
\caption{Possible Penrose diagrams 
in odd dimensions for $v_f=v_{AH}$,
which are globally naked.}
\label{fig:pd4}
\end{figure}

\begin{figure}[h]
\centerline{
\includegraphics[width=9cm]{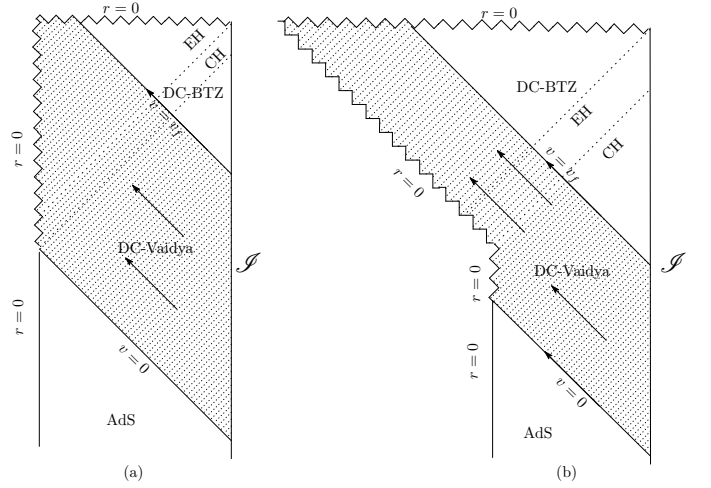}
}
\caption{Possible Penrose diagrams 
in odd dimensions with $v_f>v_{AH}$
for the globally naked singularity formation.
}
\label{fig:pd3}
\end{figure}

In the case of $0<v_f<v_{AH}$, the Penrose diagram of the gravitational collapse 
is shown in Fig. \ref{fig:oddpd}.
The possible Penrose diagrams
are depicted in  Fig. \ref{fig:pd4} for $v_f=v_{AH}$
and in Fig. \ref{fig:pd3} for $v_f>v_{AH}$.
From the present analysis, however, it is not clear whether there exists a null portion of 
the naked singularity at $r=0$ for $v=v_{AH}$.


\section{Strength of naked singularities}
\label{sec:strength}
As have seen in the preceding section,
naked singularities can be formed in DC gravity.
In this section, we investigate the 
strength of naked singularities
along the radial null geodesics with the 
asymptotic form (\ref{eq:asyeven})
in even dimensions and (\ref{eq:asyodd}) 
in odd dimensions.
We define
\begin{align}
\psi \equiv R_{\mu \nu }k^\mu k^\nu ,
\end{align}
where $k^\mu=dx^\mu /d\lambda $ is the affinely 
parametrized tangent vector of the future-directed 
outgoing radial null geodesic with an affine parameter 
$\lambda $. We evaluate the strength of naked singularities
by the divergent behavior of $\psi $
in the neighborhood of the singularities.

We consider the radial null geodesic with
the tangent vector $k^{\mu}$,
which emanates from a singularity at $\lambda =0$.
In four dimensions, the strong curvature 
condition and the limiting
focusing condition are satisfied along an
affinely parametrized geodesic
if $\lim _{\lambda \to 0}\lambda ^2\psi > 0$
and $\lim _{\lambda \to 0}\lambda \psi > 0$,
respectively
~\cite{clarke,krolak}.
However, these results have not been extended 
to the higher-dimensional case so far.
%

Along the radial null geodesics emanating 
from the singularity, we have
\begin{align}
k^r=\frac{1}{2}fk^v.
\end{align}
Then the straightforward calculation
yields
\begin{align}
 \psi =-\frac{2(D-2)\dot{f}}{rf^2}(k^r)^2,
\end{align}
where $f$ is given by Eq. (\ref{eq:fvr}).
The geodesic equation for the radial null geodesic
is written as
\begin{align}
\frac{dk^r}{d\lambda}-
\frac{2\dot{f}}{f^2}(k^r)^2=0.
\label{eq:geo}
\end{align}

As we will see below, 
the asymptotic form of the differential
equation (\ref{eq:geo}) for the DC-Vaidya solution
reduces to the following form:
\begin{align}
\frac{dk^r}{d\lambda}+sr^\alpha 
(k^r)^2\simeq 0,
\label{eq:diffeq}
\end{align}
where $s (>0)$ and $\alpha $ are constants.  
Then $\psi $ is expressed as
\begin{align}
\psi \simeq s(D-2)r^{\alpha -1}(k^r)^2.
\end{align}
We suppose the asymptotic solution of 
Eq. (\ref{eq:diffeq}) around $\lambda =\lambda _0$ as
\begin{align}
 r(\lambda )\simeq D_1(\lambda -\lambda _0)^\beta ,
\label{eq:powersol1}
\end{align}
where $D_1$ and $\beta $ are positive constants. 
The divergent behavior of $\psi$ is completely specified 
by the following two cases.\\\\
%
{\it Case 1} ) $\beta \ne 1$.
Direct substitution of Eq. (\ref{eq:powersol1})
into Eq. (\ref{eq:diffeq}) yields 
$\alpha =-1$ and $\beta =1/(1+s)$.
Redefinition of the affine parameter 
enables us to set $D_1=1$ and
$\lambda _0=0$ without loss of generality
in such a way that $\lambda =0$ corresponds to
the central singularity. 
Then we have 
\begin{align}
r(\lambda )\simeq \lambda ^{1/(1+s)}.
\end{align}
Therefore we obtain 
\begin{align}
\psi \simeq \frac{s(D-2)}{(1+s)^2}\lambda ^{-2},
\end{align}
and then it is concluded
\begin{align}
\lim _{\lambda \to 0}\lambda ^{2}\psi
=\frac{s(D-2)}{(1+s)^2}>0.
\label{eq:psi1}
\end{align}
Note that the right-hand side of Eq. 
(\ref{eq:psi1}) varies delicately with 
the scaling of the affine parameter.
\\\\
%
{\it Case 2} )  $\beta =1$.
We have to take into account the next leading term
such as 
\begin{align}
r(\lambda )\simeq D_2(\lambda -\lambda _0)
+D_3(\lambda -\lambda _0)^{1+\gamma },
\end{align}
where $\gamma $ is a positive constant.
The most dominant order of Eq. (\ref{eq:diffeq})
gives $\gamma =\alpha +1$ and 
$D_3=-sD_2^{\alpha +2}/[(\alpha +1)(\alpha +2)]$.
We set $\lambda _0=0$ and $D_2=1$
by the redefinition of the affine parameter.
Therefore the asymptotic solution is given by
\begin{align}
r(\lambda)\simeq \lambda .
\end{align}
Hence we have
\begin{align}
\psi \simeq s(D-2)\lambda ^{\alpha -1},
\end{align}
it is then concluded
\begin{align}
\lim _{\lambda \to 0}\lambda ^{1-\alpha }
\psi =s(D-2)>0.
\end{align}

Discussions for the odd- and even-dimensional cases
are made separately in the following subsection
by the aid of the criterion obtained above.

\subsection{Even dimensions}
For $q=1$, we find that the asymptotic form of Eq.
(\ref{eq:geo}) near $v=r=0$ becomes
\begin{align}
\frac{dk^r}{d\lambda }+\frac{s_1}
{r}(k^r)^2\simeq 0
\label{eq:asygeoevenq=1}
\end{align}
with
\begin{align}
s_1\equiv \frac{K_1-2}
{D-2}>0,
\end{align}
where the constant $K_1$ satisfies the 
equation (\ref{eq:keven}). 
This belongs to Case 1. We then obtain
\begin{align}
\lim_{\lambda \to 0}\lambda ^2 \psi
=\frac{s_1(D-2)}{(1+s_1)^2}>0.
\label{eq:evenpsiq=1}
\end{align}

We next examine the case where $q>1$. 
In this case, the asymptotic form of the 
geodesic equation becomes
\begin{align}
\frac{dk^r}{d\lambda }+{s_2}r^{(q-n)/(n-1)} 
(k^r)^2\simeq 0
\end{align}
with  
\begin{align}
s_2=\frac{q(2^{q+1}M_0)^{1/(n-1)}}{n-1}>0.
\end{align}
This belongs to Case 2 with
$\alpha =-1+(q-1)/(n-1)$.
Thus we obtain
\begin{align}
\lim_{\lambda \to 0}\lambda ^{2(1-(q-1)/(D-2))}
\psi =s_2(D-2)>0.
\label{eq:evenpsiq}
\end{align}
The power of $\lambda $ depends on the 
power of the mass function and the number of 
the spacetime dimensions.
It takes values from $0$ for $q=D-1$ to $2$ for $q=1$.
The divergent behavior of $\psi $
is summarized in Table \ref{table:even}.

\begin{table}[h]
\begin{center}
\caption{The divergent behavior of $\psi $
around $v=r=0$ along the radial null geodesics with
the asymptotic form (\ref{eq:asyeven}) in
even dimensions.}
\label{table:even}
\begin{tabular}{c|c|c|c|c}
\hline
$q$ & $q=1$ & $1<q<D-1$ & $q=D-1$ & 
$q>D-1$\\ \hline \hline
$\psi$ & $\lambda ^{-2}$ & 
$\lambda ^{-2(1-(q-1)/(D-2))}$ & const. 
& 0\\
\hline
\end{tabular}
\end{center}
\end{table} 

\subsection{Odd dimensions}
\label{subsection:strengthodd}
The steps in the odd dimensions
are quite similar to those in even dimensions.
We will first consider the radial 
null geodesics from the singularity 
$r=0$ and $0<v<v_{AH}$
with the asymptotic form (\ref{eq:asyodd}).
The asymptotic form of the geodesic equation 
(\ref{eq:geo}) is given by
\begin{align}
\frac{dk^r}{d\lambda}
+s_3(k^r)^2
\simeq 0,
\label{eq:geoodd}
\end{align}
where the constant $s_3$ is defined by
\begin{align}
s_3 =\frac{2q(M_0v_0^q)^{1/(n-1)}}
{(n-1)v_0(1-(M_0v_0^q)^{1/(n-1)})^2}>0.
\end{align}
This corresponds to Case 2 with $\alpha =0$.
%
Consequently, we find
\begin{align}
\lim _{\lambda \to 0}\lambda \psi =
s_3(D-2)
\label{eq:psiodd1}
\end{align}
for the singularity $r=0$ and $0<v<v_{AH}$.
It is emphasized that the strength is independent both of 
the spacetime dimensions and the power of the 
mass function.

We finally consider the geodesics from the 
singularity $v=r=0$.
The asymptotic form of the radial null 
geodesics (\ref{eq:geo}) becomes
\begin{align}
 \frac{dk^r}{d\lambda }+
{s_4}r^{-1+q/(n-1)}(k^r)^2
\simeq 0,
\label{eq:oddgeo}
\end{align}
with
\begin{align}
 s_4=\frac{q(2^qM_0)^{1/(n-1)}}{n-1}>0.
\end{align}
This corresponds to Case 2 with $\alpha =-1+q/(n-1)$.
Thus, we obtain
\begin{align}
\lim _{\lambda \to 0}\lambda ^{2(1-q/(D-1))}
\psi = s_4(D-2)
\label{eq:psiodd2}
\end{align}
for $v=r=0$.
The power of $\lambda$ takes the value from 
$0$ for $q=D-1$ to $2(1-1/(D-1))$ for $q=1$.
In particular, the strength of the singularity for 
$r=0$ and $0<v<v_{AH}$ is weaker than that 
for $v=r=0$ in the case of $(D-1)/2< q< D-1$,
while in the case of $q=(D-1)/2$, 
their strengths are the same.
The divergent behavior of $\psi $
around $v=r=0$
is summarized in Table \ref{table:odd}.

\begin{table}[h]
\begin{center}
\caption{The divergent behavior of $\psi $
at $v=r=0$ along the radial null geodesics with
the asymptotic form (\ref{eq:asyodd}) in
odd dimensions. 
It is noted that, for the naked singularity 
with $r=0$ and $0<v<v_{AH}$, $\psi \propto \lambda ^{-1}$ 
is satisfied independent of $D$ and $q$.}
\label{table:odd}
\begin{tabular}{c|c|c|c|c}
\hline
$q$ & $q=1$ & $1<q<D-1$ & $q=D-1$ 
& $q>D-1$ \\ \hline \hline
$\psi$ & $\lambda ^{-2(1-1/(D-1))}$ & 
$\lambda ^{-2(1-q/(D-1))}$ & const. 
&0 \\
\hline
\end{tabular}
\end{center}
\end{table} 


\section{Conclusions and Discussions}
\label{sec:conclusions}
In this paper, we analyzed the $D(\ge 3)$-dimensional gravitational
collapse of a null dust fluid in DC gravity.
We found an exact solution with the topology of 
$\ma M \approx \ma M^2\times \ma K^{D-2}$, 
which describes the gravitational collapse.
Applying this solution with spherical symmetry to the situation 
in which a null dust fluid radially injects into initially 
$D$-dimensional AdS spacetime, we investigated the effects of 
the Lovelock terms on the final fate of gravitational collapse.
Our model is an example of dynamical and
inhomogeneous collapse in Lovelock gravity and 
should provide better insight into the CCH in the context of 
higher-curvature gravity theories.
We supposed that (i) the power-law mass function $M(v)=M_0v^q$, 
where $M_0>0$ and $q\ge 1$, and (ii) the null geodesics obey 
a power law near the singularity.

We found that globally naked singularities can be formed in DC gravity.
Furthermore, the final states of the gravitational 
collapse differ substantially depending on whether 
the spacetime dimensions are odd or even.
This property is seen in the homogeneous collapse 
of a dust fluid in DC gravity~\cite{Ilha,comment2}.
It is also noted that odd-dimensional DC gravity is 
an exceptional case in the thin-shell collapse investigated 
in~\cite{cdcs2004}, only in which a naked singularity can emerge.

A massless ingoing null naked singularity can appear
in the even-dimensional case $(D=2n)$ for $1\leq q<D-1$. 
In the odd-dimensional case $(D=2n-1)$, on the other hand, 
a massive timelike naked singularity
is formed for any $q (\ge 1)$.
These naked singularities can be globally naked.
In three dimensions, the singularity is not 
a curvature singularity, 
where the scalar curvatures are the same as those 
for the three-dimensional AdS spacetime although 
the energy density of a null dust fluid diverges.
As a result, the formation of a naked singularity
cannot be avoided in DC gravity,
nor in general relativity.


In DC gravity, massive timelike singularities
appear in odd dimensions.
When $M(v)$ is intact in Eq.~(\ref{eq:AHodd}), 
it is easily shown that this property 
is independent of the form of $M(v)$  
as long as ${\dot M}>0$ and $M(0)=0$.
The formation of a massive timelike singularity in odd dimensions 
is considered to be a characteristic feature in Lovelock gravity.
Our results are consistent with those in general relativity 
and in Gauss-Bonnet gravity.
In general relativity, a massive timelike singularity appears 
in three dimensions, where the Einstein-Hilbert term becomes 
first nontrivial. (See Appendix \ref{sec:gr}.)
While in Gauss-Bonnet gravity, it appears in five dimensions, 
where the Gauss-Bonnet term becomes first nontrivial~\cite{hideki}.

In Sec. \ref{sec:strength}, we investigated
the strength of naked singularities.
We compare the strength by the divergent behavior of 
$\psi \equiv R_{\mu \nu }k^\mu k^\nu $.
The strength of the naked singularity at $v=r=0$ depends on 
$D$ and $q$ both in odd and even dimensions 
(see Tables~\ref{table:even} and \ref{table:odd}).
In contrast, around the massive timelike singularity at 
$r=0$ and $0<v<v_{AH}$ in odd dimensions, 
$\psi$ diverges as $\lambda ^{-1}$, independent of $D$ or $q$.
Such a divergent behavior for the 
massive timelike singularity can be seen 
as well both in general relativity in three dimensions
(see Appendix~\ref{sec:gr}) 
and in Gauss-Bonnet gravity in five dimensions \cite{hideki}. 
Massive timelike singularities 
diverging as $\lambda ^{-1}$ might be
salient features in Lovelock gravity.

We have shown in this paper that naked 
singularity formation is generic for the
power-law mass function in DC gravity.
We can speculate from the results 
that naked singularity formation will also 
be inevitable in full Lovelock gravity.
It is also extrapolated that if massive timelike 
naked singularities appear in full Lovelock gravity,
they would be developed only in odd dimensions and 
have a portion in which $\psi$ would diverge as $\lambda ^{-1}$ regardless
of the spacetime dimensions or the form of 
the mass function.
Additional investigations are required in order to clarify these points.


\section*{Acknowledgments}
We would like to thank K.~Maeda for continuous encouragement.
We would also like to thank T.~Torii and U.~Miyamoto
for fruitful discussions 
and useful comments.
This work was partially supported by the Waseda University Grant 
for The 21st Century COE Program 
(Holistic Research and Education Center for Physics 
Self-organization Systems) at Waseda University.


\appendix


\section{Geodesics from singularities}
\label{sec:geodesics}


\subsection{Existence and uniqueness}
\label{sec:CMP}
We apply the fixed-point method for proving
the existence and uniqueness of
future-directed outgoing radial null geodesics
from singularities with asymptotic form
(\ref{eq:asyeven}) and (\ref{eq:asyodd}).

We define new coordinates $\theta$ and $\chi$ as
\begin{align}
\theta &\equiv \frac{r}{v-v_0},\\
\chi &\equiv (v-v_0)^s,
\end{align}
where $s$ is a positive constant and we only consider the region with $v
\ge v_0$.
Then the geodesic equation (\ref{eq:null}) becomes
\begin{align}
\frac{d\theta }{d\chi}+\frac{1}{s\chi}(\theta -\eta )
=\eta \Psi(\theta,\chi),
\label{eq:geodesictheta}
\end{align}
with
\begin{align}
\Psi (\theta, \chi) =\frac{1}{s\chi } \left[\left(\frac{1}{2\eta }-1\right)+
\frac{\theta ^2\chi^{2/s}}{2 \eta l^2}-\frac{\Phi(\theta,\chi)}{2\eta}
\right],
\end{align}
where a new parameter $\eta$ $(0<\eta <\infty )$ has been introduced.
The form of $\Phi$ is given below for each case.

If we choose the parameter $\eta$ and $s$ to be $\eta_0$ and $s_0$,
respectively,
with which $\Psi$ is at least $C^1$ in
$\chi \ge 0, \theta>0$, then we can apply
the contraction mapping principle to
Eq.~(\ref{eq:geodesictheta}) to find that
there exists the solution satisfying $\theta(0)=\eta_0$,
and moreover that it is the unique solution of
Eq.~(\ref{eq:geodesictheta}) which is continuous at $\chi=0$.
(See~\cite{CMP,CMP2} for the proof.)

\subsubsection{even-dimensional DC-Vaidya solution}
For the even-dimensional DC-Vaidya solution,
we have $v_0=0$ and
\begin{align}
\Phi &=\left(\frac{2M_0v^{q-1}}{\theta }\right)^{1/(n-1)},\\
&=\left(\frac{2M_0\chi^{(q-1)/s}}{\theta }\right)^{1/(n-1)}.
\end{align}
For $q>1$, if we choose $\eta=\eta_0=1/2$ and $s=s_0$ satisfying
\begin{align}
0<s_0&\leq 1,\\
0<s_0&\leq \frac{(q-1)}{2(n-1)},
\end{align}
$\Psi $ becomes at least $C^1$ in $\chi\geq 0, \theta >0$, and
consequently the existence of the solution with the asymptotic form 
(\ref{eq:asyeven}) is shown.
Unfortunately, this method cannot be applied to the case with $q=1$.

\subsubsection{odd-dimensional DC-Vaidya solution}
For the odd-dimensional DC-Vaidya solution,
we have
\begin{align}
\Phi &=(M_0v^q)^{1/(n-1)},\\
&=[M_0(v_0+\chi^{1/s})^q]^{1/(n-1)}.
\end{align}
For $0<v_0<v_{AH}$, if we choose $\eta=\eta_0=(1-(M_0v_0^q)^{1/(n-1)})/
2$ and $s=s_0$ satisfying $0<s_0 \leq 1$, $\Psi $ becomes at least $C^1$
in $\chi\geq 0, \theta >0$.
For $v_0=0$, on the other hand, if we choose $\eta=\eta_0=1/2$ and $s=s_
0$ satisfying
\begin{align}
0<s_0&\leq 1,\\
0<s_0&\leq \frac{q}{2(n-1)},
\end{align}
$\Psi $ becomes at least $C^1$ in $\chi\geq 0, \theta >0$.
Consequently, the existence of the solution with the asymptotic form 
(\ref{eq:asyodd}) is shown.

\subsubsection{Vaidya-AdS solution}
We also study the general relativistic case.
For the $D$-dimensional Vaidya-AdS solution in general relativity
(see Appendix \ref{sec:gr}), we have
\begin{align}
\Phi &=\frac{M_0v^{q-D+3}}{\theta ^{D-3}},\\
&=\frac{M_0(v_0+\chi^{1/s})^{q-D+3}}{\theta ^{D-3}}.
\end{align}

First we consider the case in $D\geq 4$ dimensions, 
in which we have $v_0=0$.
For $q>D-3$, if we choose $\eta=\eta_0=1/2$ and $s=s_0$ satisfying
\begin{align}
0<s_0&\leq 1,\\
0<s_0&\leq \frac{q-(D-3)}{2},
\end{align}
$\Psi $ becomes at least $C^1$ in $\chi\geq 0, \theta >0$, and
consequently the existence of the solution with the asymptotic form 
(\ref{nullray3}) is shown.
Unfortunately, this method cannot be applied to the case with $q=D-3$.

Next we consider the three-dimensional case.
For $0<v_0<v_{AH}$, if we choose $\eta=\eta_0=(1-M_0v_0^q)/2$ and $s=s_
0$ satisfying $0<s_0 \leq 1$, $\Psi $ becomes at least $C^1$ in 
$\chi \geq 0, \theta >0$.
For $v_0=0$, on the other hand, if we choose $\eta=\eta_0=1/2$ and 
$s=s_ 0$ satisfying
\begin{align}
0<s_0&\leq 1,\\
0<s_0&\leq \frac{q}{2},
\end{align}
$\Psi $ becomes at least $C^1$ in $\chi\geq 0, \theta >0$

\subsection{On the null and causal geodesics}
\label{sec:nullgeodesics}

In Sec. \ref{sec:NSF}, we have analyzed only future-directed
outgoing radial null geodesics. 
The contraposition of the following theorem implies that 
it is sufficient to consider only the future-directed 
outgoing radial null geodesics in order to determine whether 
or not the singularity is naked.

We will prove the following theorem: 
{\it If a future-directed causal (excluding radial null)
geodesic emanates from the central singularity, 
then a future-directed radial null geodesic 
emanates from the central singularity.} 
The proof is similar to the four-dimensional case in \cite{nmg2002}.

We consider the DC-Vaidya solution, of which metric 
is given by (\ref{eq:sol}) with $k=1$.
Without loss of generality, we set the geodesics 
on the several equatorial plane thanks to the spherical symmetry.
In the DC-Vaidya spacetime, 
the tangent to a causal geodesic satisfies
\begin{eqnarray}
-f\left(\frac{dv}{d\lambda}\right)^2+2\frac{dv}
{d\lambda}\frac{dr}{d\lambda}+\frac{L^2}{r^2}=\epsilon, 
\label{hamilton}
\end{eqnarray}  
where $\lambda$ is an affine parameter, $L^2$ is 
the sum of the square of conserved angular momenta 
and $\epsilon=0,-1$ for null 
and timelike geodesics, respectively.
Then, at any point on such a geodesic,
\begin{eqnarray}
f\left(\frac{dv}{d\lambda}\right)^2 \ge 
2\frac{dv}{d\lambda}\frac{dr}{d\lambda}
\end{eqnarray}
with equality holding only for radial null geodesics.
For the future-directed outgoing geodesics, this gives
\begin{eqnarray}
\frac{dv}{dr}\ge \frac{2}{f}>0
\end{eqnarray}
and 
\begin{eqnarray}
\frac{dv}{dr}\le \frac{2}{f}<0
\end{eqnarray}
in the untrapped and trapped regions, respectively.
Therefore, 
\begin{eqnarray}
\frac{dv_{\rm CG}}{dr}>\frac{dv_{\rm RNG}}{dr}>0 \label{geodesics1}
\end{eqnarray}
and 
\begin{eqnarray}
\frac{dv_{\rm CG}}{dr}<\frac{dv_{\rm RNG}}{dr}<0 \label{geodesics2}
\end{eqnarray}
are satisfied in the untrapped and trapped regions, respectively, 
where the subscripts represent causal (excluding radial null) 
geodesics and outgoing radial null geodesics, respectively.

Let us consider the $(r,v)$-plane.
The singularity is located at $r=0$ for $v \ge 0$.
First we consider the singularity with $f<0$.
By Eq.~(\ref{geodesics2}), the past-directed ingoing geodesics 
emanating from an event with $r>0$ in the trapped region cannot 
reach the singularity at $r=0$.
Therefore, there is no future-directed outgoing geodesic 
emanating from the singularity in the trapped region, 
i.e., the singularity with $f<0$ is censored.

Next we focus on the singularity with $0\le v \le v_{\rm AH}$, 
where $f\ge 0$ holds.
Now suppose that $v=v_{\rm CG}(r)$ extends back to 
a central singularity located at $(r,v)=(0,v_{s})$, 
where $v_{s}$ satisfies $0 \le v_{s} \le v_{\rm AH}$.
There exists a portion of $v=v_{\rm CG}(r)$ which is in the untrapped region.
Let $p$ be any point on such a potion of $v=v_{\rm CG}(r)$ 
and to the future of the singularity.
Applying inequality (\ref{geodesics1}) at $p$, 
we see that $v=v_{\rm RNG}(r)$ through $p$ crosses 
$v=v_{\rm CG}(r)$ from above and hence the points on 
$v=v_{\rm RNG}(r)$ prior to $p$ must lie to the future 
of points on $v=v_{\rm CG}(r)$ prior to $p$, 
in the sense $v_{\rm RNG}(r)>v_{\rm CG}(r)$ for $r \in (0,r_{\ast})$,
where $r_{\ast}$ corresponds to $p$.
Thus, the radial null geodesics must extend back to 
$r=0$ with $v=v_0$ satisfying $v_s \le v_{0} \le v_{\rm AH}$, 
and so must emerge from the singularity.

It is noted that $v_{\rm AH}=v_s=v_{0}=0$ 
holds in the even-dimensional case.

\section{Curvature Tensors}
\label{sec:sc}
We present the scalar curvatures
in order to analyze the singularities.
The curvature invariants for the 
metric (\ref{eq:sol}) are
\begin{widetext}
\begin{align}
\ma I_1&\equiv R_{\alpha \beta \gamma \delta }
R^{\alpha \beta \gamma \delta }=
(f'')^2+\frac{2(D-2)}{r^2}(f')^2
+\frac{2(D-2)(D-3)}{r^4}(k-f)^2, \\
\ma I_2&\equiv R_{\alpha \beta }
R^{\alpha \beta  }=\frac{1}{2}(f'')^2+\frac{D-2}{r}f'f''
+\frac{D(D-2)}{2r^2}(f')^2-\frac{2(D-2)(D-3)}{r^3}(k-f)f'
+\frac{(D-2)(D-3)^2}{r^4}(k-f)^2.
\end{align}
\end{widetext}
The Ricci scalar becomes
\begin{align}
R=-f''-\frac{2(D-2)}{r}f'+\frac{(D-2)(D-3)}{r^2}(k-f).
\end{align}
These formulae are applicable 
not only for the DC-Vaidya solution but also for 
the DC-BTZ solution (\ref{eq:exsol}) or 
the Vaidya-AdS solution (\ref{f-eq}).
These quantities remain finite, as expected, 
for the three-dimensional BTZ solution~\cite{BTZ}.

\section{General relativistic collapse}
\label{sec:gr}

In general relativity, the function $f(v, r)$ 
in Eq.~(\ref{eq:sol}) and the energy density 
$\rho$ for the $D$-dimensional 
Vaidya-AdS solution for $k=1$ are
\begin{equation}
\label{f-eq}
f(v, r)=1-\frac{M(v)}{r^{D-3}}+l^{-2}r^2
\end{equation}  
and 
\begin{eqnarray}
\rho=\frac{D-2}{2\kappa_D^2r^{D-2}}{\dot M}, 
\label{density}
\end{eqnarray}  
respectively, where $M(v)$ is an arbitrary function of $v$.
$\kappa _D^2$ is related to the $D$-dimensional gravitational
constant $G_D$ as $\kappa _D^2=8\pi G_D$.
We assume the form of $M(v)$ as a power law
\begin{eqnarray}
M(v)=M_0 v^q \label{massform}
\end{eqnarray}  
as that in the case of DC gravity, where 
$M_0$($>0$) and $q (\geq 1)$ are constants.
We can see the central singularity appears at $r=0$
for $v>0$. The singular nature at $v=r=0$ will be
clarified below.
The trapped region is
\begin{eqnarray}
M(v)=M_0 v^q\geq r^{D-3}+l^{-2}r^{D-1}. \label{ah}
\end{eqnarray}
We will consider the $D=3$ and $D\geq 4$ cases separately
in the following.

\subsection{$D\geq 4$ case}
From Eq. (\ref{ah}),
we can find that only the point $v=r=0$ 
has the possibility of being naked.
Thus, a massive timelike naked singularity is absent in this case 
for any mass function with $M(0)=0$ and $\dot M\geq 0$.
%

We assume that the asymptotic form of 
the future-directed outgoing null geodesics 
near $v=r=0$ as
\begin{eqnarray}
v \simeq K_3 r^{p},\label{nullray0}
\end{eqnarray}
where $K_3$ and $p$ are positive constants.
After some straightforward calculations, 
we find the asymptotic solution
\begin{eqnarray}
v \simeq 2r
\label{nullray3}
\end{eqnarray}
for $q>D-3$.
For $q=D-3$, which is valid for $D>3$, we obtain 
\begin{eqnarray}
v \simeq K_3r,
\label{nullray4}
\end{eqnarray}
where $K_3$ satisfies the relation
\begin{align}
M_0K_3^{D-2}-K_3+2=0.
\label{eq:K3} 
\end{align}
The condition for the existence for 
positive $K_3$ satisfying Eq. (\ref{eq:K3})
is given by 
\begin{align}
0<M_0\leq \frac{1}{D-2}\left(\frac{D-3}{2(D-2)}
\right)^{D-3}.
\label{eq:masgr}
\end{align}
When $M_0$ satisfies Eq. (\ref{eq:masgr})
for $q=D-3$,
the spacetime represents the formation of
the naked singularities; otherwise
we need a more careful investigation
to determine the final state of the 
gravitational collapse. 
The right-hand side of Eq. (\ref{eq:masgr})
goes to zero as $D\to+\infty $; 
here the naked singularity 
formation becomes less feasible
as the spacetime dimensions are higher,
similar to the case of the even-dimensional 
DC-Vaidya solution.
While, there exist no null geodesics with
the asymptotic form (\ref{nullray0}) for
$1\leq q<D-3$.
The geodesics (\ref{nullray3}) and (\ref{nullray4}) are singular 
null geodesics for $D-3\le q<D-1$.
Thus, the solution represents the formation of a massless 
ingoing null naked singularity for $D-3<q<D-1$ for any 
$M_0(>0)$ and $q=D-3$ with $M_0$ satisfying Eq.~(\ref{eq:masgr}).

The asymptotic form of the null geodesic 
equation near $v=r=0$ is given by
\begin{eqnarray}
\frac{d}{d\lambda}k^r +\frac{s_5}{r}(k^r)^2
\simeq 0\label{beqgr}
\end{eqnarray}
with
\begin{eqnarray}
s_5 \equiv \dfrac{{K_3^{D-2}M_0(D-3)}}{2}
\end{eqnarray}
for $q=D-3$, where $K_3$ satisfies the relation 
(\ref{eq:K3}). This belongs to Case 1
in Sec. \ref{sec:strength}.
For $q>D-3$, 
The asymptotic form of the null geodesic 
equation near $v=r=0$ is given by 
\begin{align}
\frac{d}{d\lambda}k^r +{s_6}r^{q-D+2}(k^r)^2
\simeq 0
\end{align}
with
\begin{align}
s_6\equiv 2^qM_0q.
\end{align}
This belongs to Case 2 with $\alpha =q-D+2$.

It is then concluded
\begin{align}
\lim _{\lambda \to 0}\lambda ^2\psi
=\frac{s_5(D-2)}{(1+s_5)^2}
\end{align}
for $q=D-3$, and
\begin{align}
\lim _{\lambda \to 0}\lambda ^{D-(q+1)}\psi
=s_6(D-2) 
\end{align}
for $D-3<q<D-1$.
The divergent behavior of $\psi $
is summarized in Table \ref{table:gr}.

\begin{table}[h]
\begin{center}
\caption{The divergent behavior of $\psi $
around $v=r=0$ along the radial null geodesics with
the asymptotic form (\ref{nullray3}) for $D>q-3$
and (\ref{nullray4}) for $q=D-3$ in $D\geq 4$ dimensions.
}
\label{table:gr}
\begin{tabular}{c|c|c|c|c}
\hline
$q$ & $q=D-3$ & $D-3<q<D-1$ & $q=D-1$ &
$q>D-1$ \\ \hline \hline
$\psi$ & $\lambda ^{-2}$ & 
$\lambda ^{-D+q+1}$ & const. 
& 0\\
\hline
\end{tabular}
\end{center}
\end{table} 

\subsection{$D=3$ case}
From Eq. (\ref{ah}),
the point $r=0, 0\leq v\leq v_{AH}=M_0^{-q}$ can be naked.
In this case, the discussion in Sec. \ref{subsec:odd}
can be applied. 
There exist null geodesics which behave as 
Eq. (\ref{eq:asyodd}) with $n=2$ for $0\leq v< v_{AH}$.
The scalar curvature quantities are finite everywhere
along these null geodesics
although the energy density of the null dust fluid diverges.
From Sec. \ref{subsection:strengthodd}, 
along the null geodesics from $v=r=0$,
$\psi =R_{\mu \nu }k^\mu k^\nu $ diverges as 
$\lambda ^{-2+q}$.
In contrast, along the null geodesics from $r=0$, $0<v<v_{AH}$,
$\psi$ diverges as $\lambda ^{-1}$ independent of the power of 
the mass function.



\begin{thebibliography}{99}
\bibitem{comment2}
The singularities shown in the Penrose diagrams 
in~\cite{Ilha} are spacelike both in odd and even dimensions.
However, the singularity in the dust region 
is actually null in odd dimensions 
because the scale factor depends on the cosmic time 
$t$ linearly in the vicinity of $t=0$, 
while it is spacelike because the scale factor 
in even dimensions behaves as 
$a(t)\propto t^{1-1/(D-1)}$ near $t=0$.
\end{thebibliography}
\end{document}